\documentclass[twocolumn,prx,superscriptaddress,floatfix]{revtex4-1}

\usepackage{amsmath,amssymb,mathrsfs}
\usepackage{natbib}
\usepackage{tabularx}
\usepackage{amsfonts}
\usepackage{subcaption}
\usepackage{amsmath}
\usepackage{comment}
\usepackage{bbold} 
\usepackage{hhline}
\usepackage{braket}
\usepackage{txfonts}
\usepackage{balance}
\usepackage{physics}
\usepackage{graphicx}
\usepackage{dcolumn}
\usepackage{bm}
\usepackage[utf8]{inputenc}
\usepackage[english]{babel}
\usepackage[T1]{fontenc}
\usepackage{mathtools}

\usepackage[unicode=true,bookmarks=true,bookmarksnumbered=false,bookmarksopen=false,breaklinks=false,pdfborder={0 0 1},backref=false,colorlinks=true,citecolor=blue,linkcolor=blue]{hyperref}

\def\be{\begin{equation}}
\def\ee{\end{equation}}
\def\bea{\begin{eqnarray}}
\def\eea{\end{eqnarray}}

\newcommand{\RNum}[1]{\uppercase\expandafter{\romannumeral #1\relax}}

\begin{document}

\title{Solving The Travelling Salesman Problem Using A Single Qubit}% Force line breaks with \\

\author{Kapil Goswami} 
\email{kgoswami@physnet.uni-hamburg.de}
\affiliation{%
 Zentrum f\"ur Optische Quantentechnologien, Universit\"at Hamburg, Luruper Chaussee 149, 22761 Hamburg, Germany
}%
\author{Gagan Anekonda Veereshi}
\affiliation{%
Zentrum f\"ur Optische Quantentechnologien, Universit\"at Hamburg, Luruper Chaussee 149, 22761 Hamburg, Germany
}%

\author{Peter Schmelcher}
\affiliation{%
 Zentrum f\"ur Optische Quantentechnologien, Universit\"at Hamburg, Luruper Chaussee 149, 22761 Hamburg, Germany
}%
\affiliation{%
 The Hamburg Centre for Ultrafast Imaging, Universit\"at Hamburg, Luruper Chaussee 149, 22761 Hamburg, Germany
}%

\author{Rick Mukherjee}%
\email{rmukherj@physnet.uni-hamburg.de}
\affiliation{%
 Zentrum f\"ur Optische Quantentechnologien, Universit\"at Hamburg, Luruper Chaussee 149, 22761 Hamburg, Germany
}%

\date{\today}

\begin{abstract}
The travelling salesman problem (TSP) is a popular NP-hard-combinatorial optimization problem that requires finding the optimal way for a salesman to travel through different cities once and return to the initial city. The existing methods of solving TSPs on quantum systems are either gate-based or binary variable-based encoding. Both approaches are resource-expensive in terms of the number of qubits while performing worse compared to existing classical algorithms even for small-size problems.
We present an algorithm that solves an arbitrary TSP using a single qubit by invoking the principle of \textit{quantum parallelism}.
The cities are represented as quantum states on the Bloch sphere while the preparation of superposition states allows us to traverse multiple paths at once. The underlying framework of our algorithm is a quantum version of the classical Brachistochrone approach.
Optimal control methods are employed to create a selective superposition of the quantum states to find the shortest route of a given TSP.
The numerical simulations solve a sample of four to nine cities for which exact solutions are obtained. The algorithm can be implemented on any quantum platform capable of efficiently rotating a qubit and allowing state tomography measurements. For the TSP problem sizes considered in this work, our algorithm is more resource-efficient and accurate than existing quantum algorithms with the potential for scalability. A potential speed-up of polynomial time over classical algorithms is discussed.
\end{abstract}
\maketitle

There is growing interest in developing efficient quantum algorithms to solve combinatorial optimization problems \cite{farhi2014quantum,han2002quantum,han2000genetic,farhi2019quantum,Goswami}. 
TSP is a well-known combinatorial optimization problem that is NP-hard, hence, classical polynomial-time algorithms are not known to solve them \cite{korte_2018}.
The general TSP and its variations have multiple direct mappings to routing problems and extended applications such as machine scheduling, cellular manufacturing, arc routing, frequency assignment, structuring of matrices, and DNA sequencing \cite {gutin2006traveling,lenstra}.
There exist classical algorithms to solve a general TSP \cite{arora1998polynomial,bellman1962dynamic,held1962dynamic,woeginger2003exact}, however, the time to obtain the exact solution classically scales exponentially with the number of cities. The heuristic classical algorithms do not guarantee optimal solution(s) \cite{gutin2006traveling}, nevertheless, they can run a million-city TSP \cite{applegate2003implementing} and provide approximate solutions whose optimality is difficult to verify due to the NP-hardness of the problem. Hence, there is ongoing research for developing quantum algorithms that can solve TSP to gain a practical advantage in terms of time and accuracy compared to the classical counterparts.

In many of the quantum approaches, the problem is mapped to a quadratic unconstrained binary optimization (QUBO) form and solved using the Ising Hamiltonian-based quantum annealer \cite{jain2021solving,kieu,warren2020solving}.
There are also approaches based on phase-estimation methods used in quantum circuits \cite{tszyunsi2023quantum,srinivasan2018efficient,cryptoeprint:2024/626,zhu2022realizable}, and the success of these algorithms depends on the availability of a large number of noiseless qubits, which is a bottleneck for current experiments.
Typically, current quantum algorithms have a success probability of less than $90\%$ for a 4-city TSP \cite{tszyunsi2023quantum} which decreases with increasing problem size.
The quantum algorithm \cite{jain2021solving} for encoding 9 and 10-city problems on a D-wave quantum architecture requires 73 logical qubits or 5436 physical qubits. Thus, there is a huge scope for improvement in solving TSP on a quantum system.

\begin{figure*}[t]
\includegraphics[width = 1\textwidth,trim={0cm 10cm 1cm 0cm},clip]{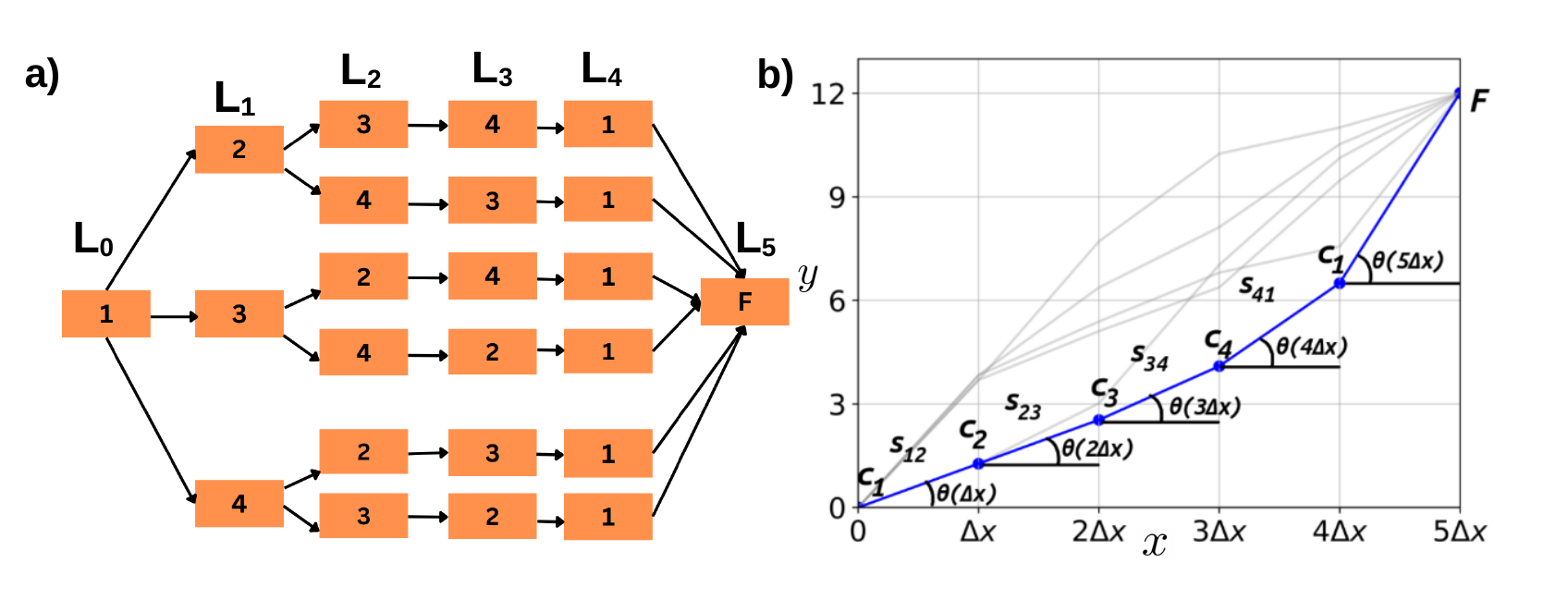}
\caption{TSP as a Brachistochrone problem. 
(a) A classical routing chart with $6$ layers (each denoted by $L_i$) and cities as orange rectangles representing all the paths for a $4$-city TSP. The last additional rectangle $F$ is needed as all the routes end at a common final point in the brachistochrone problem. 
(b) A graphical representation of a $4$-city TSP with linear piece-wise functions describing all the possible routes as solid lines. The length $s_{ij}$ of each line encodes the distance between the cities $c_i$ and $c_j$, parameterized by the angle $\theta({i\Delta x})$. Each city and piece-wise line in (b) is represented as a rectangle (or circle) and an arrow in (a), respectively. One of the routes is highlighted in blue, which is used as an example for describing the graphical construction in Subsection \ref{TSPB}.}
  \label{fig1}
\end{figure*}

In this work, we map the (a)symmetric TSP onto a single qubit and solve it by optimally driving the system to achieve a selective population distribution of the internal levels. Entanglement is often considered one of the necessary ingredients to gain a quantum advantage, however, a polynomial speedup can also be achieved solely by the superposition principle \cite{lloyd1999quantum,kenigsberg2006quantum,goswami2024integer}. In our algorithm, the TSP is mapped to a classical discrete brachistochrone problem on a plane with constraints, and the solution of the brachistochrone problem corresponds to the optimal route for the given TSP. The finite brachistochrone-TSP is mapped to a Bloch sphere with cities represented by quantum states of a single qubit and the optimal route is found by steering the qubit from one state to another in a specific manner using rotation operators. We use the superposition of states to travel through multiple paths at once and pick out the optimal one by post-selecting from the projective measurement of only the penultimate state. For each part of a path, rotation operators and their strengths are tuned by the Simultaneous Perturbation Stochastic Approximation (SPSA) algorithm \cite{SPSA,spall1998overview}. We solve 4- to 9-city TSPs using a single qubit and exploit quantum parallelism to get the optimal path (approximation ratio of $\sim 1$) for more than $90\%$ of the problem.
In the rare cases where we do not have exact solutions, high approximation ratios ($\sim 0.90$) are obtained. Any quantum platform such as superconducting qubits \cite{10.1063/1.5089550,PhysRevLett.129.010502}, trapped ions \cite{srinivas2021high,schafer2018fast}, nitrogen-vacancy centers in diamond \cite{doi:10.1126/sciadv.aar7691} and Rydberg atoms \cite{Adams_2020,PhysRevLett.121.123603,RevModPhys.82.2313}, that allows arbitrary qubit rotation with high fidelity can implement our algorithm. 

In Section \ref{theory}, we introduce the TSP and map it to a finite discrete brachistochrone problem in Subsection \ref{TSPB}. The problem is then carefully encoded on a Bloch sphere using specialized mapping which is discussed in Subsection \ref{TSPBB}. 
Optimal control methods to solve the encoded problem and a detailed discussion about the measurement scheme are provided in Subsection \ref{TSPBB}. Section \ref{results} presents the results of a brute-force as well as the quantum optimization approach using the superposition principle to solve TSP on a Bloch sphere. Finally, Section \ref{conc} contains a discussion on the potential polynomial speed-up our algorithm can have over classical algorithms as well as an outlook for future research.

\section{\label{theory}Theory}

\subsection{\label{TSP}Travelling Salesman problem}

Mathematically, there are multiple formulations of the TSP such as binary integer programming (including Miller Tucker Zemlin (MTZ)~\cite{pataki2003teaching} and the Dantzig Fulkerson Johnson (DFJ)~\cite{oncan2009comparative}), binary quadratic programming~\cite{zaman2021pyqubo} and dynamic programming~\cite{chauhan2012survey}. For our work, TSP is viewed as a graph problem whose solution corresponds to the shortest Hamiltonian cycle. The cost function of the problem can be defined to minimize the total travelling distance or the fuel cost, both of which can be encoded on the graph with weights on the edges. Generally, for an $n$-city TSP, the distances/fuel costs $s_{ij}$ between the cities $c_i$ and $c_j$ are represented by a matrix, $B=(s_{ij})_{n\times n}$, known as \textit{cost matrix}. The problem is symmetric if $s_{ij}=s_{ji}~\forall i,j$ and asymmetric if there exists $i,j$ for which $s_{ij}\neq s_{ji}$. 
There are $(n-1)!$ possible TSP routes where each one is represented by a Hamiltonian cycle $H_b$ for $b \in \{1,\hdots,(n-1)!\}$, where $b$ keeps track of different Hamiltonian cycles. An arbitrary $H_b$ is of the form $c_i \rightarrow c_j \rightarrow c_k \rightarrow \hdots \rightarrow c_i$. The cost function of TSP is then defined as,

\be
\begin{split}
 D = \min_{b} \qty(\sum_{\{c_i,c_j\} \in H_b} s_{ij}),    
\end{split}
 \label{mainobj}
\ee

where only the pair of cities occurring in the sequence of a Hamiltonian cycle $H_b$ are included in the sum. The quantity inside the bracket above is calculated for individual cycles $H_b$ with the goal of finding the shortest $H_b$. In this work, we choose the problem of minimizing the total distance while travelling through all the routes as an example of TSP.

\subsection{\label{TSPB}TSP as a finite discrete Brachistochrone problem}
The brachistochrone problem \cite{haws1995exploring} is a classical optimization problem to find the path with minimal time traveled by a particle between the initial $I$ and final $F$ points. To solve it, a time functional has to be minimized, given as,

\begin{align}
     \int_{I}^{F} \frac{ds}{v}, 
\end{align}

where $v$ is the velocity of the particle, and the above brachistochrone problem is continuous. In the discrete version of this problem, the objective is to find a linear piece-wise function that minimizes the time taken for the journey. When the number of sub-intervals tends to infinity, the path naturally converges to the continuous case \cite{agmon2020discrete}.

The discrete version of a constrained brachistochrone problem is used to encode and solve the TSP for the classical case before generalizing it to its quantum version. Consider a prototype 4-city TSP as an example for which all the routes are shown by the routing chart in Fig.~\ref{fig1}(a) with each layer $L_i$ consisting of multiple cities $c_1,c_2...c_n$. The orange rectangles represent cities $c_i$ with arrows showing connections between them. In subsequent sections, we will refer to routing charts to distinguish different routes taken during the algorithm. 
For a 2D geometric representation of the problem, consider a coordinate system consisting of $x$ and $y$ axes as shown in Fig.~\ref{fig1}(b), with the first city $c_1$ placed at the origin. The rest of the cities $c_i$ are placed on the 2D plane where the $x$-axis is divided into $(n+1)$ sub-intervals of equal length $\Delta x$ that is less than the distance between the closest cities in the problem. 
For a given TSP route, the adjacent cities $c_i$ and $c_j$ are placed such that the difference between their coordinates $(x_i-x_j,y_i-y_j)$ is $(\Delta x, \sqrt{s_{ij}^2-\Delta x^2})$ with the Euclidean distance being $s_{ij}$ as shown in Fig.~\ref{fig1}(b). All such linear paths, denoted by gray lines on a 2D plane are constructed corresponding to all the possible routes in TSP.
Traditional brachistochrone problem demands a unique final point and by definition of TSP, each route finally returns to the same initial city ($c_1$ at $x=n \Delta x$). This aspect of the problem is encoded by adding an extra linear piece defined from $c_1$ to $F$ at $x=(n+1)\Delta x$ to each route as shown in Fig.~\ref{fig1}.
The length of each linear part of a route is expressed by the angle it subtends with the $x$-axis and any TSP route $H_b$ is parametrized by a discrete function $\theta_b(x)$ for all $b\in \{0,...,(n-1)! \}$. Specifically, any two cities $c_i,c_j$ occurring successively (as $(k+1)$th and $(k+2)$th elements) in a Hamiltonian cycle $H_b = \{c_1,\hdots, c_i, c_j, \hdots,c_1\}$ are placed at $x = k\Delta x$ and $x = (k+1)\Delta x$ respectively with the distance between them given by $s_{ij} = \Delta x/cos(\theta_b(k \Delta x))$. 
At a particular $x=k\Delta x$, $\theta_b(x)$ can take multiple values corresponding to the placed cities belonging to different possible TSP routes $H_b$. The time functional to find the optimal set of $\theta_b(x)$ for domain $x \in \{0,\Delta x,...,n\Delta x\}$ describing the distances for all the paths is given as,
\begin{align}
    \begin{split}
        Q=\min_{b}~\qty(\frac{\Delta x}{v}\sum_{x=0}^{(n-1)\Delta x} \frac{1}{ cos(\theta_b(x))}),
    \end{split}
    \label{time}
\end{align}
where $v$ is the constant velocity for all the routes. The term inside the bracket is calculated for different functions $\theta_b(x)$ and a specific one provides the solution to the TSP problem. The last linear piece for all the routes from the initial city to $F$ is left out of the sum in $Q$ as it is only needed to complete the brachistochrone construction.
Minimizing the time functional $Q$ gives the optimal TSP path. In the next section, we discuss how to implement the discrete brachistochrone problem on a Bloch sphere.

\begin{figure*}[t]
\centering
\includegraphics[width = 1\textwidth,trim={6.5cm 0cm 0cm 0cm},clip]{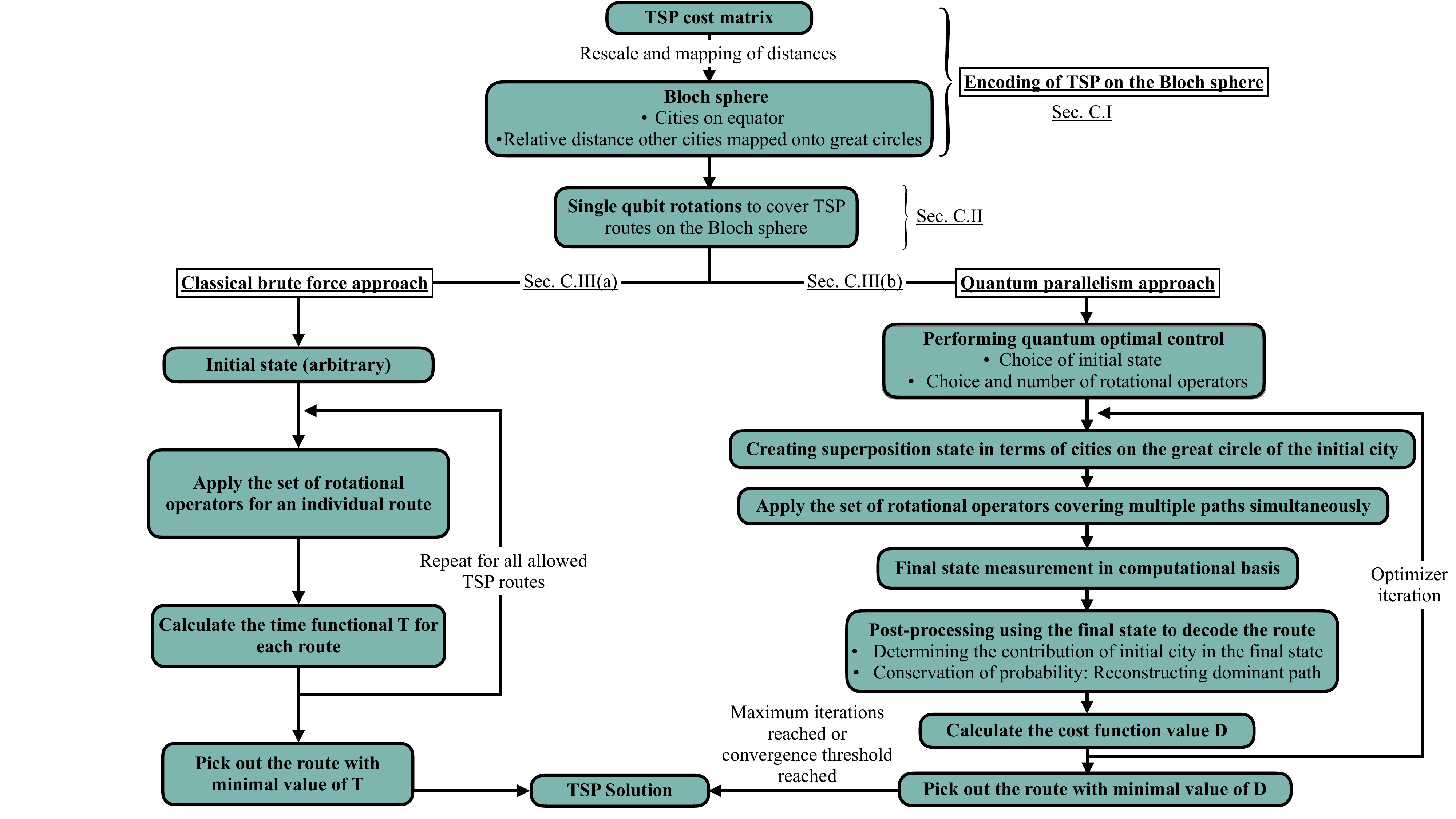}
\caption{The chart shows the main components and the corresponding sub-steps of the algorithm along with the reference to the sections in our manuscript for detailed discussion. The flow diagram is divided into three parts, where the encoding scheme (top part) is common for the classical brute force algorithm (left part) and the quantum parallelism algorithm (right part). The quantum optimal control has several hyper-parameters, such as the choice of the initial state, the choice and the number of rotation operators to be optimized, and the maximum optimizer iterations. The choice of the objective function to be minimized i.e., the time functional $T$ for the brute force approach and the cost function $D$ for the quantum approach is explained in the text. The details for the individual parts of this chart are provided in Section~C.}
  \label{fig_flow}
\end{figure*}

\begin{figure*}[t]
\centering
\includegraphics[width = 1\textwidth,trim={0cm 8cm 1cm 0cm},clip]{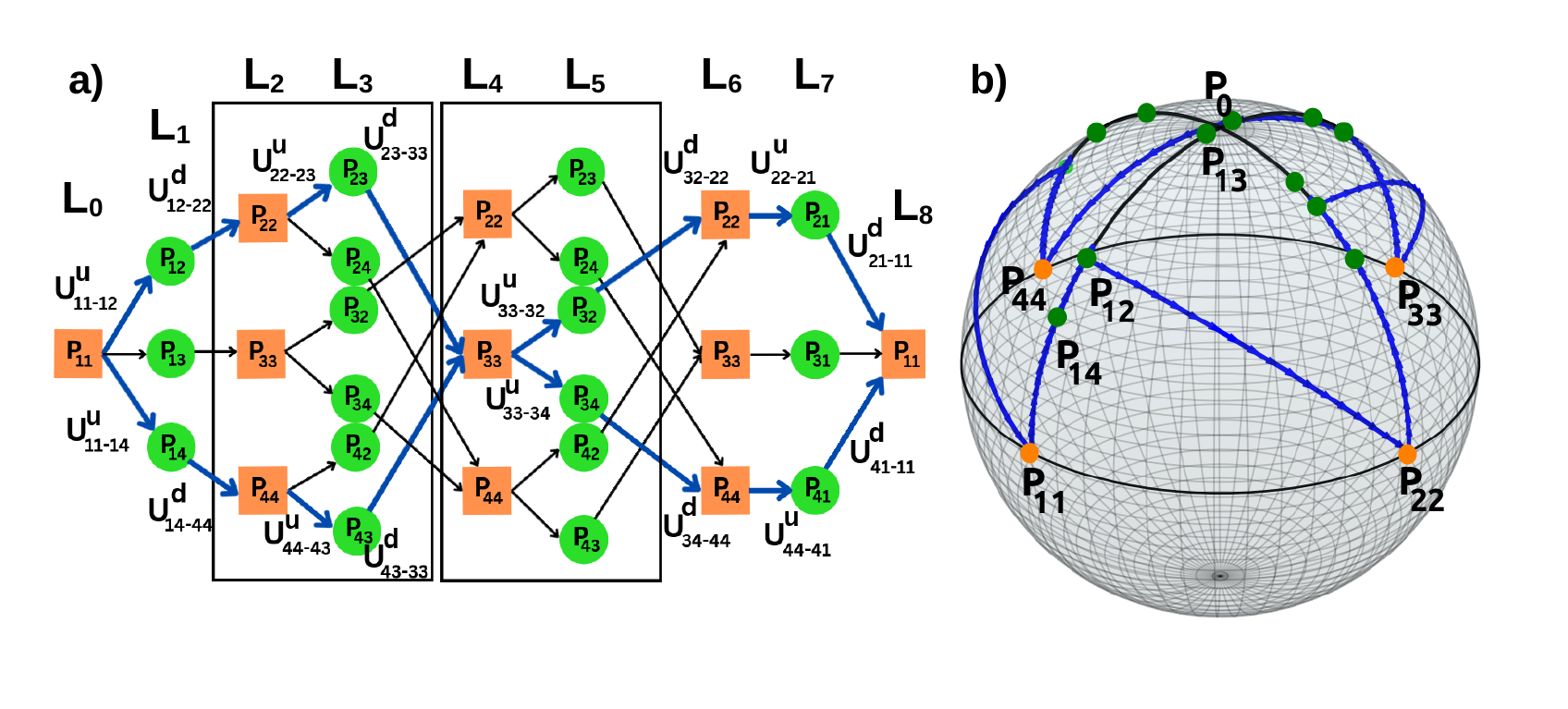}
\caption{TSP as a finite discrete Brachistochrone problem. 
(a) Quantum routing chart of all the classical routes of a general $4$-city TSP with $9$ layers. The orange rectangle and the green circles correspond to the cities in the original problem and auxiliary city-states that encode relative distances respectively. 
The routes indicated by blue arrows $P_{11}\rightarrow$ $P_{12} \rightarrow$ $ P_{22} \rightarrow$ $ P_{23} \rightarrow$ $ P_{33} \rightarrow$ $ P_{34} \rightarrow$ $ P_{44} \rightarrow$ $ P_{41} \rightarrow P_{11}$ and $P_{11}\rightarrow $ $ P_{14} \rightarrow$ $ P_{44} \rightarrow$ $ P_{43} \rightarrow$ $ P_{33} \rightarrow$ $ P_{32} \rightarrow$ $ P_{22} \rightarrow$ $ P_{21} \rightarrow P_{11}$ correspond to the optimal solutions $c_{1}\rightarrow $ $c_{2}\rightarrow $ $c_{3}\rightarrow $ $c_{4}\rightarrow c_{1}$ and $c_{1}\rightarrow$ $ c_{4}\rightarrow $ $c_{3}\rightarrow $ $c_{2}\rightarrow c_{1}$ respectively for the example problem whose cost matrix is given by Eq.~(\ref{cm1}) in appendix \ref{costm}, which are found by the algorithm discussed in Section \textbf{\RNum{3}(a)}.
(b) Schematic diagram of the optimal classical route on the Bloch sphere. The states $P_{ii}$ (orange points on the equator) represent the cities in the problem. The distance between cities $c_i$ and $c_j$ is encoded by the geometric distance between the quantum states $\ket{P_{ii}}$ and $\ket{P_{ij}}$ (green point on the geodesic connecting $P_{ii}$ and the pole $P_0$), which is explicitly shown for $\ket{P_{12}},\ket{P_{13}},\ket{P_{14}}$. 
The solid-blue route represents the optimal path, corresponding to the blue-arrow path in (a).  
To go from one state to another state on the Bloch sphere, rotation operators $U^u_{ii-ij}$ and $U^d_{ij-ii}$ are used, and, $u$,$d$ corresponds to the path going towards $P_0$ ($P_{11}$ to $P_{12}$) and away from $P_0$ ($P_{12}$ to $P_{11}$) respectively.}
  \label{fig2}
\end{figure*}

\begin{figure*}[t]
\centering
\includegraphics[width = 1.0\textwidth,trim={0cm 8cm 0cm 0cm},clip]{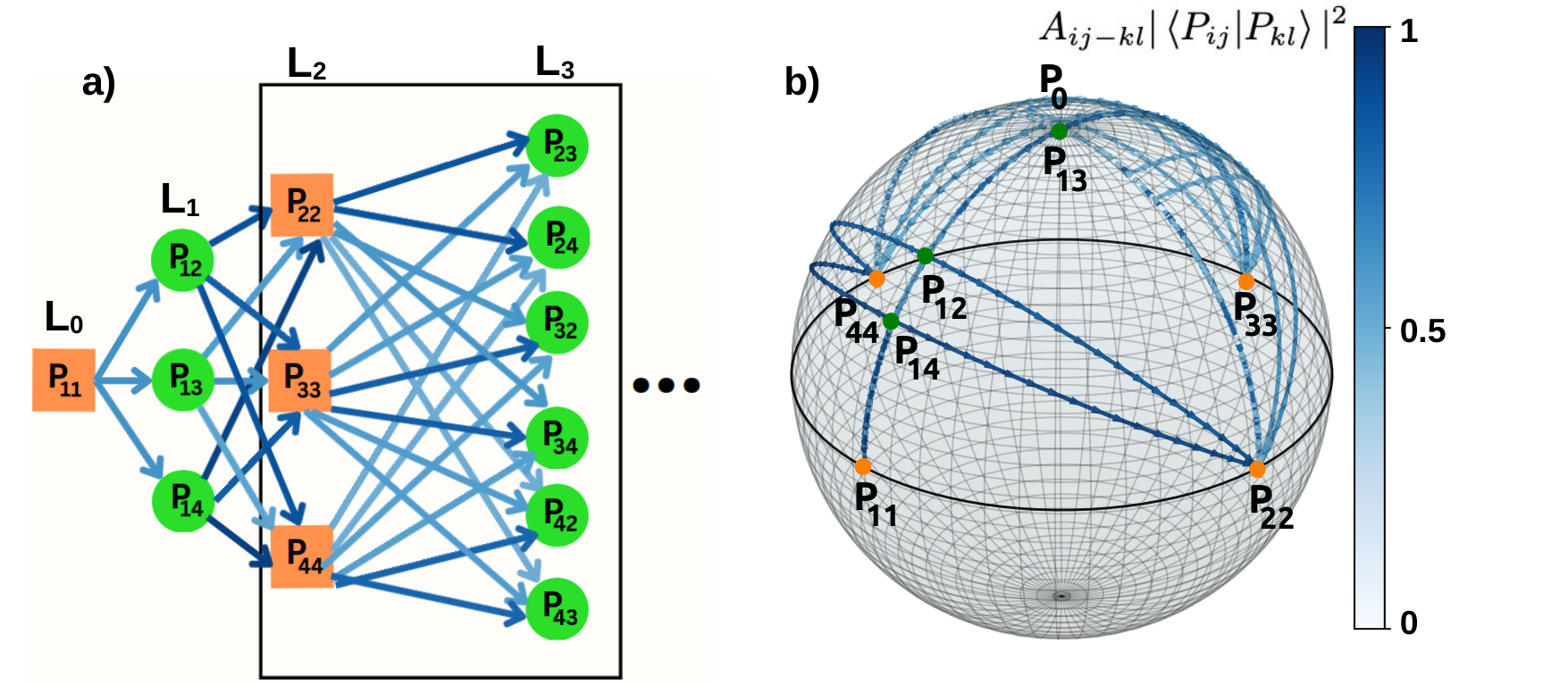}
\caption{TSP solved on the basis of the superposition principle. 
(a) The routing chart shows multiple routes for the initial four layers of a general $4$-city TSP, with a finite population transfer between the states corresponding to the non-classical paths. All the possible paths are traveled at once by the algorithm discussed in Section \textbf{\RNum{3}(b)}.
(b) Schematic diagram showing different population transfers from the quantum states of one layer to the quantum states of another layer corresponding to the optimal protocol on the Bloch sphere, the color-bar $A_{ij-kl}|\braket{P_{ij}}{P_{kl}}|^2$ quantifies the population transfer between the states.}
  \label{fig4}
\end{figure*}

\subsection{\label{TSPBB}Solving TSP-brachistochrone problem on the Bloch sphere}

In general, encoding an arbitrary TSP on a Bloch sphere is a non-trivial procedure. The cities on a 2D plane are to be represented by the quantum states such that the paths on the Bloch sphere encode corresponding TSP paths. 

Before encoding the problem to the Bloch sphere, a routing chart is constructed to visualize all the essential components required to represent a TSP.
For an example of a $4$-city problem, a routing chart consisting of nine layers $L_i,~i \in \{0,1,...,8\}$, series of vertical rectangles (and circles) corresponding to the cities and arrows connecting them to represent possible TSP routes is shown in Fig.~\ref{fig2}(a), which is schematically similar to Fig.~\ref{fig1}(a).  
In particular, the cities $c_1,c_2,c_3,c_4$ in Fig.~\ref{fig1}(a) are replaced by orange rectangles $P_{11},P_{22},P_{33},P_{44}$ with intermediate green-circles $P_{ij}$ in Fig.~\ref{fig2}(a). In contrast to Fig.~\ref{fig1}(a), the layer consisting of $P_{22},P_{33},P_{44}$ is repeated in Fig.~\ref{fig2}(a) which is essential for encoding distances on the Bloch sphere.
The orange rectangles $P_{ii}$ are used to determine a TSP route uniquely in Fig.~\ref{fig2}(a), where the arrows have a physical meaning as they represent rotation operators. 

Let the Hamiltonian cycle for a particular quantum path $b$ be denoted by $\mathcal{H}_b$. For example, a route $\mathcal{H}_1$ described as $P_{11}\rightarrow P_{12}\rightarrow P_{22}\rightarrow P_{23}\rightarrow P_{33}\rightarrow P_{34}\rightarrow P_{44}\rightarrow P_{41} \rightarrow P_{11}$ corresponds to a TSP path $H_1$ in Fig.~\ref{fig1}(a), given by $c_{1}\rightarrow c_{2}\rightarrow c_{3}\rightarrow c_{4}\rightarrow c_{1}$.
A complete overview of the algorithm is provided by Fig.~\ref{fig_flow} comprising both the encoding and the steps for solving the problem. The flow chart bifurcates with the two resulting branches (left and right) contrasting the procedure of finding the solution using the classical and quantum approaches. The key difference between the two algorithms lies in the method of accessing the different routes, i.e., separately and individually in the classical approach while simultaneously multiple routes in the quantum scheme.
The explicit steps for encoding and solving the TSP on the Bloch sphere are given below.\\

\noindent \textbf{\RNum{1}. Encoding the cities on the Bloch sphere}.
The key issue for this type of encoding is to preserve the relative distances between all the pairs of cities when translating them from a 2D plane to a sphere \cite{robinson2006sphere}. A careful mapping similar to the inverse-stereographic projection \cite{Lisle_Leyshon_2004} is required and one such mapping is given here to encode TSP onto a Bloch sphere with the sub-steps mentioned in Fig.~\ref{fig_flow}. 

Each rectangle (and circle) in the routing chart corresponds to a quantum state $\ket{P_{kl}}$ (both the states with $k=l$ and $k\neq l$) on the Bloch sphere (in Fig.~\ref{fig2}(b)). Any such state $\ket{P_{kl}}$ with Cartesian coordinates $(x_{kl},y_{kl},z_{kl})$ on a unit sphere is expressed in the computational basis as,
\begin{align}\label{eqn_psi}
    \ket{P_{kl}}=\cos{\frac{\xi_{kl}}{2}}\ket{0}+e^{i\frac{\phi_{kl}}{2}} \sin{\frac{\xi_{kl}}{2}\ket{1}}
\end{align}
with $R=1$, $\xi_{kl} = \cos^{-1}{z_{kl}}$ and $\phi_{kl} = \text{sign}(y_{kl}) \cdot \arccos(x_{kl}/\sqrt{x_{kl}^2+y_{kl}^2})$ are its spherical coordinates.
For an $n$-city problem, the states $\{\ket{P_{11}}, \ket{P_{22}}, \ket{P_{33}},..., \ket{P_{nn}}\}$ are placed on the equator of the Bloch sphere, shown as the orange points in Fig.~\ref{fig2}(b). These states divide the equator into $n$ equal parts with the initial state placed arbitrarily.
From each of the equator states $\ket{P_{ii}}$, $n$-geodesic curves are constructed by connecting them to one of the poles $P_0$.
On each of the geodesic curves ($P_{ii}-P_0$), states $\ket{P_{ij}}~\forall j\neq i$ (green points in Fig.~\ref{fig2}(b)) are placed such that the geometric distance between $\ket{P_{ii}}$ and $\ket{P_{ij}}$ on the unit sphere corresponds to the relative distance $s_{ij}$ between the cities $c_i$ and $c_j$ for all $j\neq i$.
For example, in Fig.~\ref{fig2}(b), the states $\ket{P_{12}},\ket{P_{13}},\ket{P_{14}}$ are placed on the $P_{11}-P_0$ curve.
The geometric distance from any point on the equator to the pole is $\pi/2$, so the scaled distances must be less than or equal to $\pi/2$.
For implementing our algorithm we initialize the system with the quantum state $\ket{P_{11}}$ on the Bloch sphere, also shown by the first rectangular orange box in Fig.~\ref{fig2}(a). \\

\noindent \textbf{\RNum{2}. Travelling between cities on a Bloch sphere}.
After placing the quantum states on each geodesic, the rotation operators can be used to access them on the Bloch sphere. Specifically, the rotation operators connect the states on the Bloch sphere (thick blue curves in Fig.~\ref{fig2}(b)) along a geodesic path. The general rotation operator taking any state $\ket{P_{ij}}$ to $\ket{P_{kl}}$ on the Bloch sphere is given by,
\begin{align}
        U^{u/d}_{ij-kl}=cos\frac{\delta}{2}-i sin\frac{\delta}{2} (\mathbf{n} . \mathbf{\sigma})
        \label{U}
\end{align}
where $\delta = cos^{-1}(\braket{P_{ij}}{P_{kl}})$ is the rotation angle, $\mathbf{n}=[n_1,n_2,n_3]$ is the unit vector (normal to both $\ket{P_{ij}}$ and $\ket{P_{kl}}$) defining the axis of rotation, and $\mathbf{\sigma} = [\hat{\sigma}_1,\hat{\sigma}_2,\hat{\sigma}_3]$ is a Pauli matrix-vector. Superscripts $u$ or $d$ are labels indicating \textit{up} or \textit{down} rotations corresponding to taking states towards or away from the pole $P_0$. 
The initial state $\ket{P_{11}}$ is rotated to reach the next randomly chosen state $\ket{P_{jj}}$ by two rotation operators. (1) The first rotation $U^u_{11-1j}$ takes the state $\ket{P_{11}}$ \textit{up} from the equator to state $\ket{P_{1j}}$ closer to pole $P_0$. It covers the real distance between cities $c_1$ and $c_j$ in the process. (2) The second rotation $U^d_{1j-jj}$ takes $\ket{P_{1j}}$ \textit{down} to the state $\ket{P_{jj}}$ on equator, away from $P_0$. It resets the system for applying (1) to travel to the next cities. Application of (1)-(2) takes the quantum states from one orange rectangle to another and defines a sub-route,
\begin{equation}
   P_{11} \xrightarrow{(1)U^u_{11-1j}} P_{1j} \xrightarrow{(2)U^d_{1j-jj}} P_{jj} 
\end{equation}
as shown in Fig.~\ref{fig2} (for an example case with $j=2$). It covers the path between cities $c_1 \rightarrow c_j$. Alternatively, projection operators of the form $\ket{P_{1j}}\bra{P_{11}}$ and $\ket{P_{jj}}\bra{P_{1j}}$ can be used respectively, to go from one state to another.

\noindent \textbf{\RNum{3}(a). Solving of TSP on a Bloch sphere using a brute force approach}.
The two-step rotation is applied such that the already covered quantum states are left out when choosing the next state, ensuring each city is visited once. For example, in the path
\begin{equation}
\begin{split}
\begin{aligned}
    P_{11}\xrightarrow{U^u_{11-1j}} P_{1j}\xrightarrow{U^d_{1j-jj}} P_{jj} \xrightarrow{U^u_{jj-jk}} P_{jk} \\    \xrightarrow{U^d_{jk-kk}} P_{kk}, 
\end{aligned}
\end{split}
\end{equation}
city $j$ is covered, so all the next operations are forbidden to go to $\ket{P_{kj}}$. The process is repeated until the initial state is reached again after covering all the others exactly once as shown in Fig.~\ref{fig2}.
In this way, one of the possible $\mathcal{H}_b$ on the Bloch sphere is travelled every time the arrows are followed from the first layer to the last in the routing chart. In total, $2n-1$ rotations should be done to cover one possible TSP path for an $n$-city problem. The next step is to find the optimal route out of all the possibilities.
For all the TSP paths $\mathcal{H}_b$ with $b\in \{ 0,...,(n-1)!\}$ on the Bloch sphere, the total time functional $T$ is given as,
\begin{align}
\begin{split}
    T= \min_{b}~\qty(\sum_{P_{ii},P_{jj} \in \mathcal{H}_b} \tau^b_{ii-ij}),
\end{split}
    \label{T}
\end{align}
where $\tau^b_{ii-ij}$ is the time for the system going from an arbitrary state $\ket{P_{ii}}$ to another $\ket{P_{ij}}$ along a particular $\mathcal{H}_b$. $\tau^b_{ii-ij}$ is the overlap of the states for the \textit{up} sub-routes of the form $\ket{P_{ii}}\rightarrow \ket{P_{ij}}$ (going from city $c_i$ to city $c_j$) and defined as optimal time \cite{carlini2006time}, given by, 
\begin{align}
    \tau^b_{ii-ij}=\frac{1}{|{\omega}|} cos^{-1}\braket{P_{ij}}{P_{ii}}, 
    \label{tau}
\end{align}
with $\omega$ being a non-zero constant depending on the problem (for more details see \cite{carlini2006time}). $\omega$ is set to $1$ in our case, without loss of generality. $\tau^b_{ii-ij}=0$ when the states $\ket{P_{ii}}$ and $\ket{P_{ij}}$ are the same, and it is maximum when the states are orthonormal (diametrically opposite on the Bloch sphere), indicating that it captures the geodesic distance between any two points on the sphere.  There is a one-to-one correspondence between the classical time functional $Q$ from Eq.~(\ref{time}) and time functional $T$ from Eq.~(\ref{T}) on the Bloch sphere which is numerically shown in the results. $T$ is the total time required to cover a given TSP path, and the optimal route corresponds to the smallest value of $T$, indicated by the bold-blue path in Fig~\ref{fig2}. 
Each path is traversed one at a time and thus is still a brute-force approach but on a Bloch sphere. The left part of Fig~.\ref{fig_flow} concisely show the steps for this brute force approach. The quantum generalization of this protocol will be discussed in the next section.\\

\noindent \textbf{\RNum{3}(b). Solving the TSP on a Bloch sphere using the superposition principle}.
This section describes the components and procedure for solving the TSP using the quantum superposition principle on the Bloch sphere. Initially, a routing chart is constructed to visually represent multiple paths traversed simultaneously by applying rotation operators on the Bloch sphere. We decompose the resulting superposition state into pre-encoded quantum states on the Bloch sphere to gain a better understanding. The objective is to steer the quantum system to a decomposition of the penultimate superposition state at the $2n$th layer of the algorithm into the already placed quantum states whose population contribution encodes a valid classical Hamiltonian cycle. The Hamiltonian cycle is used to calculate the classical cost function $D$ (as defined in Eq.~(\ref{mainobj})) where the problem is solved by identifying the optimal rotation operators that minimize $D$. The penultimate state can be measured by performing quantum state tomography. A complete run of the algorithm contains creating multiple superposition states on the Bloch sphere, initializing the chosen optimizing parameters, and a final measurement for post-processing to determine the dominant path in that run. The parameters are then iteratively optimized to minimize $D$. These aspects will now be discussed in detail. The steps of the quantum version of the algorithm are briefed in the right part of Fig.~\ref{fig_flow}. These steps are explained by taking a $4$-city TSP as an example which is schematically depicted in Fig.~\ref{fig4}.

\noindent \textit{Creation of the superposition states to access multiple TSP paths at once:} The routing chart in Figs.~\ref{fig4}(a) is similar to Fig.~\ref{fig2}(a), consisting of the same rectangles (and circles) and layers but distinguishing itself in terms of the initialization scheme and consequently the paths that follow.
Specifically, the quantum states at each layer $L_i$ of the routing chart in Fig.~\ref{fig4}(a) create a superposition state $\ket{g_{i}}$ that helps the system to traverse multiple TSP routes simultaneously by applying rotation operators on the Bloch sphere ($U^{u}$ and $U^d$). The first state $\ket{g_{0}} = \ket{P_{11}}$ is chosen arbitrarily as shown in Fig.~\ref{fig4}(a,b). The system is initialized with a superposition state $\ket{g_1}$ at layer $L_1$ using the operator $\alpha_{12}U^u_{11-12} + \alpha_{13}U^u_{11-13}+\alpha_{14}U^u_{11-14}$ which acts on $\ket{g_{0}}$ as shown,
\begin{equation}
\begin{split}
   \begin{aligned}
   \MoveEqLeft \ket{g_{1}} \\
   &= (\alpha_{11-12}U^u_{11-12} + \alpha_{11-13}U^u_{11-13} \\
    &\quad +\alpha_{11-14}U^u_{11-14})
   \ket{P_{11}} \\
    &= \alpha_{11-12}U^u_{11-12} \ket{P_{11}}+ \alpha_{11-13}U^u_{11-13}\ket{P_{11}} \\
    &\quad  +\alpha_{11-14}U^u_{11-14}\ket{P_{11}} \\
     &= \alpha_{11-12}\ket{P_{12}} + \alpha_{11-13}\ket{P_{13}} + \alpha_{11-14}\ket{P_{14}}
   \end{aligned}
\end{split}
\label{g1}
\end{equation}
where $\alpha_{ii-ij} \in \mathbb{C}$ is the coefficient associated with the rotation operator $U^u_{ii-ij}$.
The state $\ket{g_{1}}$ is a single point when represented on the Bloch sphere and can be decomposed into the quantum states ($\ket{P_{12}},\ket{P_{13}},\ket{P_{14}}$) of layer $L_1$. In general, this decomposition is non-unique and, the coefficients $\alpha_{ii-ij}$ associated with the rotation operators govern one of the ways to quantify the contribution of these states as shown by Eq.~(\ref{g1}). A second set of operators $\alpha_{12-22} U^d_{12-22}$, $\alpha_{13-33} U^d_{13-33}$ and $\alpha_{14-44} U^d_{14-44}$ are applied with the goal to rotate \textit{down} each component of $\ket{g_{1}}$ (without the coefficients in front) respectively. The overall state at layer $L_2$ state is given by $\ket{g_{2}} = (\alpha_{12-22} U^d_{12-22} + \alpha_{13-33} U^d_{13-33} + \alpha_{14-44} U^d_{14-44}) \ket{g_{1}}$. Each of these \textit{down} operators e.g. $U^d_{12-22}$ acts on all the components of $\ket{g_1}$, rotating them to reach the state $\ket{P_{22}}$ as well as the states $\ket{P^{j=3,4}_{22}}$.
To obtain a better insight into the contributions of $\ket{P^{j=3,4}_{22}}$, we look at the first term in $\ket{g_{2}}$ which is expanded below,
\begin{equation}
\begin{split}
\begin{aligned}
  \MoveEqLeft \alpha_{12-22} U^d_{12-22}\ket{g_{1}} \\
    &= \alpha_{12-22} U^d_{12-22}(\alpha_{11-12}\ket{P_{12}} + \alpha_{11-13}\ket{P_{13}}  \\
                &\quad + \alpha_{11-14}\ket{P_{14}})  \\
                 &= \alpha_{12-22} (\alpha_{11-12} \ket{P_{22}}+ \alpha_{11-13}\ket{P^{3}_{22}}  \\
                &\quad + \alpha_{11-14}\ket{P^{4}_{22}})
\end{aligned}
\end{split}
\label{g12}
\end{equation}
where $\ket{P_{22}^{j=3,4}}$ represent the states that deviate from the intended final state $\ket{P_{22}}$, and $\alpha_{ij-jj}$ describes the prefactor of the $U^d_{ij-jj}$ operator. $U^d_{12-22}$ is shown as the arrow connecting the states $\ket{P_{12}}$ to $\ket{P_{22}}$ in Fig.~\ref{fig4}(a). The states $\ket{P_{22}^{j=3,4}}$ lie on the geodesic connecting $P_{22}$ to $P_0$, given as,   
\begin{equation}
\begin{split}
\begin{aligned}
    \ket{P_{22}^{j}} = \cos{\frac{(\xi_{22}+\xi_{1j}-\xi_{12})}{2}}\ket{0}+ \\ e^{i\frac{\phi_{22}}{2}} \sin{\frac{(\xi_{22}+\xi_{1j}-\xi_{12})}{2}\ket{1}},
\end{aligned}
\end{split}
\end{equation}
where, $(\xi_{12},\phi_{12})$, $(\xi_{22},\phi_{22})$ and $(\xi_{1j},\phi_{22})$ are the coordinates of the states $\ket{P_{12}}$, $\ket{P_{22}}$ and $\ket{P_{1j}}$ on the Bloch sphere. For $j=2$ the above state $\ket{P^j_{22}}$ is $\ket{P_{22}}$. In general, the states of the form $\ket{P_{kk}^{j}}$ arise due to the non-orthogonality of the $\ket{P_{ik}}$ and $\ket{P_{ij}}$ from the previous layer.
The state $\ket{g_2}$ is also represented by a single point on the Bloch sphere which is decomposed into $\ket{P_{22}},\ket{P_{33}},\ket{P_{44}}$ at layer $L_2$. The states $\ket{P_{22}},\ket{P_{33}},\ket{P_{44}}$ on the equator are non-parallel and hence form an \textit{over-complete} basis that can be used to describe any state on the Bloch sphere. The contribution of the states $\ket{P_{ii}}$ to $\ket{g_2}$ will be a non-trivial superposition of the states $\ket{P_{ii}}$ and $\ket{P_{ii}^{j}}$ with complex coefficients, which is determined by the cumulative effect of the previous layers in the algorithm. 
The blue trajectories with varying opacities in Fig.~\ref{fig4}(a,b) schematically show the population transfer between the quantum states when traversing from one layer to another on the Bloch sphere.
In this way, each state at layer $L_{i+1}$ receives non-zero contributions from all the quantum states in the previous layer $L_{i}$ as depicted by the arrows in Fig.~\ref{fig4}(a).

\begin{figure*}[t]
\centering
\includegraphics[scale=0.55,trim={0cm 12cm 0cm 0cm},clip]{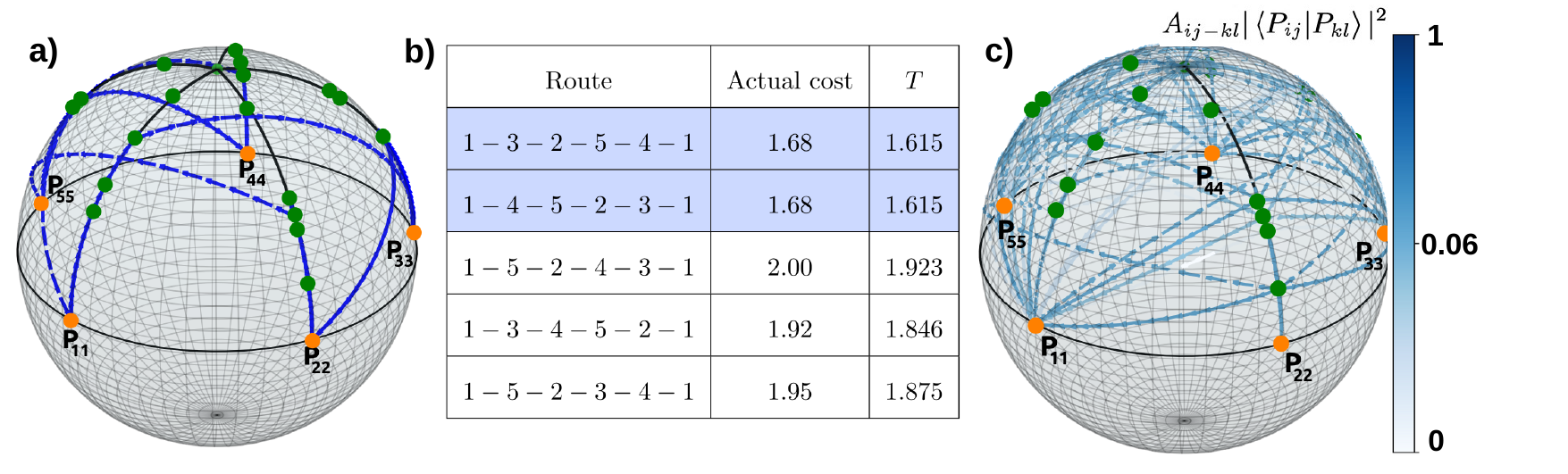}
\caption{Result for a 5-city symmetric-TSP problem whose cost function is given by the matrix in Eq.~(\ref{cm1}). (a) The route $P_{11}\rightarrow$ $P_{12} \rightarrow$ $ P_{22} \rightarrow$ $ P_{23} \rightarrow$ $ P_{33} \rightarrow$ $ P_{34} \rightarrow$ $ P_{44} \rightarrow$ $ P_{41} \rightarrow P_{11}$ corresponding to the classical optimal TSP path $c_{1}\rightarrow $ $c_{2}\rightarrow $ $c_{3}$ $\rightarrow c_{4}\rightarrow c_{1}$ is found using the brute-force algorithm of Section \textbf{\RNum{3}}(a). (b) The table shows the one-to-one correspondence of the TSP cost $D$ and time $T$ for the brute-force approach, where the optimal solution is highlighted. It shows 5 out of 24 paths. (c) All the simultaneous routes are shown that are found by the quantum approach discussed in Section \textbf{\RNum{3}}(b) for the optimized parameters. The colored paths show the population transfer, quantified by $A_{ij-kl}|\braket{P_{ij}}{P_{kl}}|^2$, between the quantum states $\ket{P_{ij}}$ and $\ket{P_{kl}}$ connected on the Bloch sphere. Due to the non-orthogonality of the quantum states on the Bloch sphere, there is a finite population transfer between the states that are not part of a valid Hamiltonian cycle $\mathcal{H}_b$, this aspect is also elaborated in Section \textbf{\RNum{3}}(b).}
\label{fig3a}
\end{figure*}

\noindent \textit{Identifying optimizing parameters and the cost function to be minimized:} 
Unlike the brute force approach, $T$ defined in Eq.~(\ref{T}) cannot be used as the objective function for this algorithm. Superposition states created at each layer in the quantum case also cover the paths that are not valid Hamiltonian cycles (or classical TSP paths), which is indicated by the additional paths in the routing chart given in Fig.~\ref{fig4}(a) as compared to the one in Fig.~\ref{fig2}(a). The exact number and scaling of these classically forbidden paths with the number of cities $n$ is given by Eq.~\ref{forbnum} with the calculation in Appendix~\ref{calc}. Optimizing for $T$ in the quantum case will include these classically \textit{unwanted} paths, thus minimizing or maximizing it solves a completely different problem rather than TSP. To identify and sample through exclusively the valid Hamiltonian cycles, we resort back to the classical cost function $D$ (Eq.~(\ref{mainobj})) that needs to be minimized. The contributions of the states $\ket{P_{j1}}$, $j\neq 1$ to the penultimate superposition state encodes the information required to sample valid Hamiltonian cycles and calculate the corresponding cost function $D$. These contributions to the penultimate state depend on the rotation operators throughout the algorithm, which act as the optimizing parameters to guide the system to the optimal Hamiltonian cycle. 

For an $n$-city problem, there are a total of $2(n-1)(2+(n-2)^2)$ (Eq.~\ref{opernum}) independent rotational operators ($36$ for $n=4$, $88$ for $n=5$), and simultaneous tuning all of them (both $U^u$ and $U^d$) to reach a particular interference, can be a complex task. The calculation for the number of rotational operators is given in Appendix~\ref{rotnum}. An issue with having too many varying $U^{u}$ and $U^{d}$ is that many \textit{undesired} paths (non-Hamiltonian cycles) are accessed and excessive suppression results in sub-optimal solutions due to a limited number of pathways. 
Hence, we only vary a few \textit{up} operators $U^u$ where the choice of the number and which of the $U^u$ to be varied are typical hyper-parameters for our algorithm that depend on the specific TSP to be solved.   
Specifically, for an $n$-city problem, more than $n$ intermediate operators (same as the number of layers where states are rotated \textit{up}) are selected to be varied. This ensures that at least one rotation operator is controlled when traversing from layer $L_{i-1}$ to $L_{i+1}$. 
We now address the specifics of the optimization process, including the measurement scheme and the post-selection of data necessary for evaluating $D$. 

\noindent \textit{Measurement and post-processing to optimize the cost function:}
For a $n$-city problem, there are $2n+1$ layers in the algorithm (as shown in Fig.~\ref{fig2}), however, we focus on the penultimate $2n$th layer at which the system reaches a superposition state $\ket{g_{2n-1}}$. 
Using quantum state tomography methods \cite{schmied2016quantum}, the state $\ket{g_{2n-1}}$ on the Bloch sphere can be estimated experimentally by measuring observables such as $\expval{\hat{\sigma}_x}{g_{2n-1}}$, $\expval{\hat{\sigma}_y}{g_{2n-1}}$ and $\expval{\hat{\sigma}_z}{g_{2n-1}}$.
Now we detail the post-processing of the experimental data. The measured state $\ket{g_{2n-1}}$ is first expressed in the computational basis (such as Eq.~(\ref{eqn_psi})). All the TSP paths return to the same initial city by definition, which in the case of the Bloch sphere are encoded by the states $\{\ket{P_{21}}...\ket{P_{n1}}\}$ forming the $2n$th layer. To find the dominant Hamiltonian cycle traversed by the system, $\ket{g_{2n-1}}$ is decomposed into the quantum states $\{\ket{P_{21}}...\ket{P_{n1}}\}$ and then the corresponding contribution of these states are calculated. Specifically, the state is represented as $\ket{g_{2n-1}}= \beta_{21}\ket{P_{21}} + \beta_{31}\ket{P_{31}}+ ... + \beta_{n1}\ket{P_{n1}}$ with the coefficients $\beta_{j1}$ still need to be determined. The experimental data for $\ket{g_{2n-1}}$ and known quantum states $\{\ket{P_{21}}...\ket{P_{n1}}\}$ are used explicitly to calculate the overlaps $\braket{P_{j1}}{g_{2n-1}}$ for all $j \in \{2,3,...,n\}$. It gives rise to the following set of $n-1$ linear equations represented by the matrices to find the unknown coefficients $\beta_{j1}$ as
\begin{gather}
    \mathcal{E} \mathbf{X} = \mathbf{K}, 
    \label{EXK}
\end{gather}

\begin{gather}
\begin{split}
\begin{aligned}
 \MoveEqLeft \mathcal{E} = \\ &\begin{bmatrix}
   1 &  \braket{P_{21}}{P_{31}} & \hdots &  \braket{P_{21}}{P_{n1}} \\
   \braket{P_{31}}{P_{21}} & 1  & \hdots &  \braket{P_{31}}{P_{n1}} \\
   \vdots &  \vdots &  &  \vdots \\
   \braket{P_{n1}}{P_{21}} &  \braket{P_{n1}}{P_{31}} & \hdots &  1
   \end{bmatrix}_{n-1\cross n-1} \\
\mathbf{X} &=    
   \begin{bmatrix}
   \beta_{21} \\ \beta_{31} \\ \vdots \\ \beta_{n1}
   \end{bmatrix}_{n-1\cross 1} ,  
   \mathbf{K} = 
 \begin{bmatrix} \braket{P_{21}}{g_{2n-1}} \\ \braket{P_{31}}{g_{2n-1}} \\ \vdots \\ \braket{P_{n1}}{g_{2n-1}} \end{bmatrix}_{n-1\cross 1}
 \end{aligned}
 \end{split}
  \label{MEQ}
\end{gather}

In general, all of the quantum states $\{\ket{P_{21}}...\ket{P_{n1}}\}$ lie on different geodesics forming a non-orthogonal \textit{overcomplete} basis \cite{duffin1952class} if $n > 3$ and a \textit{complete} basis for $n=3$ (TSP problem is trivial for $n<3$). Hence, representing $\ket{g_{2n-1}}$ in terms of $\ket{P_{j1}}$ captures its full dimensionality without any loss of information. This dimensionality can be reduced for certain special cases of the problem when the above $\mathcal{E}$ matrix is singular. For example, if any two quantum states coincide on the Bloch sphere, two specific rows or columns are the same in $\mathcal{E}$, making it singular. In such a case two elements of the vectors $\mathbf{K}$ and $\mathbf{X}$ also become equal, reducing the dimensionality of the problem from $n-1$ to $n-2$. Finding the coefficient vector $\mathbf{X}$ stays deterministic even in this special case. 

The contribution of the states $\ket{P_{j1}}$ to $\ket{g_{2n-1}}$ consists of two components: the known part, $|\braket{P_{j1}}{g_{2n-1}}|^2$, which is determined through measurement, and the unknown part, represented by the coefficients $|\beta_{j1}|^2$, which are carried forward through the algorithmic steps.
The dominant Hamiltonian cycle taken by the system is determined by arranging the cities $c_j$ in descending order of the contributions $|\beta_{j1}|^2, j \in \{2,...,n\}$ of the quantum states to $\ket{g_{2n-1}}$. This is due to the conservation of probability at each layer, which is later confirmed with our numerical results.
The order of the cities at the penultimate layer $2n$ is then used to calculate the cost function $D$ (Eq.~(\ref{mainobj})) that needs to be minimized. Different methods, such as gradient, non-gradient, Bayesian \cite{PhysRevLett.125.203603,mukherjee2020preparation}, and stochastic algorithms can carry out the optimization of the cost function $D$. The state $\ket{g_{2n-1}}$ is then rotated back to the initial state at $(2n+1)^{th}$ layer to complete the brachistochrone construction. 

\begin{figure*}[t]
\centering
\includegraphics[scale=0.55,trim={0cm 12cm 0cm 0cm},clip]{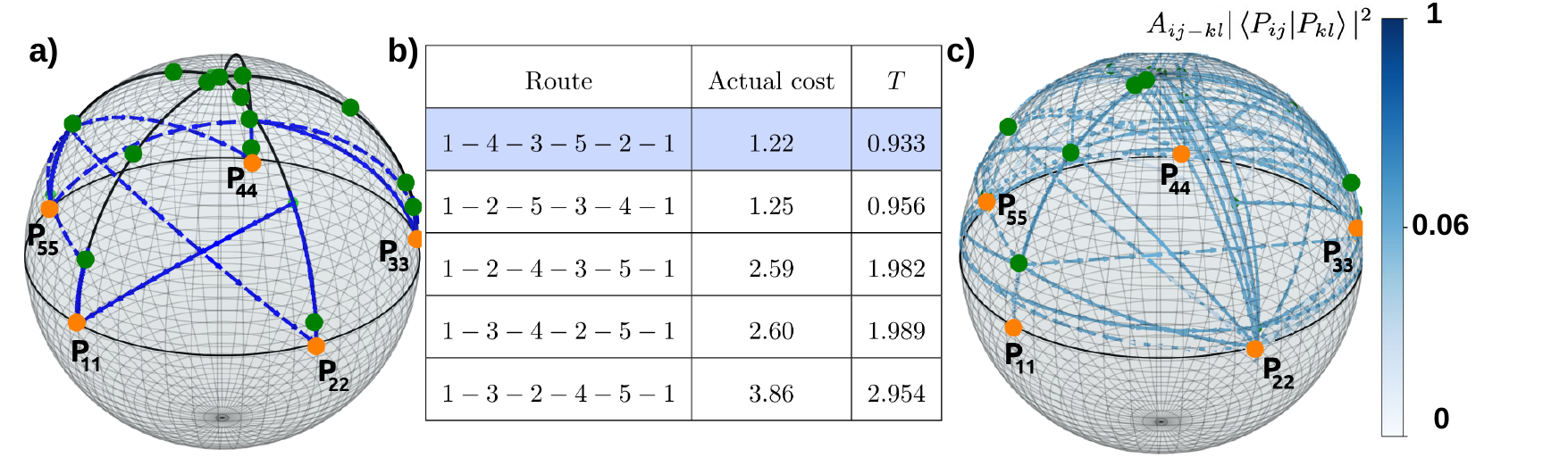}
\caption{Result for a 5-city asymmetric-TSP problem whose cost matrix is given by Eq.~(\ref{cm2}). (a) The route $P_{11}\rightarrow$ $P_{14} \rightarrow$ $ P_{44} \rightarrow$ $ P_{43} \rightarrow$ $ P_{33} \rightarrow$ $ P_{32} \rightarrow$ $ P_{22} \rightarrow$ $ P_{21} \rightarrow P_{11}$ corresponding to the classical optimal TSP path $c_{1}\rightarrow c_{4}\rightarrow c_{3}\rightarrow c_{2}\rightarrow c_{1}$ is found using the brute-force algorithm of Section \textbf{\RNum{3}}(a). (b) The table shows the one-to-one correspondence of the TSP cost and time $T$, with the optimal solution highlighted. It shows 5 out of 24 paths. (c) All the simultaneous routes are shown that are found by the quantum approach discussed in Section \textbf{\RNum{3}}(b) for the optimized parameters. The colored paths show the population transfer, quantified by $A_{ij-kl}|\braket{P_{ij}}{P_{kl}}|^2$, between the quantum states $\ket{P_{ij}}$ and $\ket{P_{kl}}$ connected on the Bloch sphere.}
\label{fig3b}
\end{figure*}

Let us illustrate the above procedure at hand with a specific example. We take a symmetric $4$-city problem with a cost matrix given by Eq.~(\ref{cm1}) in appendix \ref{costm}.
$\ket{P_{11}}$ is chosen as the initial state which corresponds to the city $c_1$ and the matrix $\mathcal{E}$ is obtained by calculating the overlaps using the quantum states defined by the encoding as $\braket{P_{j1}}{P_{k1}}$ for $j,k \in \{2,3,4\}$. The vector $\mathbf{K}$ contains the information of the state measured at the penultimate layer with elements of the form $\braket{P_{j1}}{g_{7}},~j \in \{2,3,4\}$. The coefficient vector $\mathbf{X}$ containing $\beta_{j1}$ is found by solving Eq.~(\ref{EXK}) and the corresponding contributions $|\beta_{j1}|^2$ are calculated as $|\beta_{21}|^2 = 0.62$, $|\beta_{31}|^2 = 0.23$ and $|\beta_{41}|^2 = 0.15$. These contributions are arranged in descending order, where the first subscript provides the encoded cities, $c_2,c_3,c_4$. It translates to the shortest Hamiltonian cycle  $c_1 \rightarrow c_2 \rightarrow c_3 \rightarrow c_4 \rightarrow c_1$ which is the solution to the given TSP.

The dominant platform-dependent experimental factors that result in sub-optimal solutions are imprecise measurement of the quantum state, and errors accumulating by imperfect application of rotation operators. Distinguishing any two \textit{closeby} states on the Bloch sphere is quantified by precision $\epsilon$ of the measurements which contributes to the robustness of the predictive power of the experiment. Determining the expectation values $\expval{\hat{\sigma}_x}$, $\expval{\hat{\sigma}_y}$, $\expval{\hat{\sigma}_z}$ with a precision $\epsilon$ requires $1/\epsilon^2$ experimental measurements for each observable i.e., for $\epsilon=0.01$ radians, a total of $30,000$ measurements are needed for three observables. The time required for each measurement is also an important experimental factor that depends on the platform used: a single measurement on a superconducting qubit takes $1-10 \mu s$ \cite{walter2017rapid}; on a trapped ion qubit, it takes $\sim 10 \mu s$ \cite{fluhmann2019encoding} while a photonic qubit takes $\sim 10 ns$ \cite{PhysRevLett.123.250503}. For example, in a superconducting qubit platform, the total time needed for $30,000$ measurements including experimental factors such as preparing and resetting for each measurement amounts to $\sim 3s$. An error analysis with imperfect rotations is discussed in the results section.

\section{\label{results}Results}
This section presents the results from the classical and quantum implementation of the algorithm on a Bloch sphere to solve $5$- to $8$-city prototypical symmetric and asymmetric TSPs.

Figs.~\ref{fig3a}(a)~and~\ref{fig3b}(a) show the optimal routes for prototypical symmetric and asymmetric 5-city TSPs respectively (the corresponding cost matrices are given by the Eqs.~(\ref{cm1})~and~(\ref{cm2}) in the appendix \ref{costm}), found by the classical algorithm implemented on the Bloch sphere.
Both cases have 24 possible routes, with a two-fold degeneracy for the symmetric case. The actual costs $D$ (Eq.~(\ref{mainobj})) and $T$ (Eq.~(\ref{T})) of all these routes have a one-to-one correspondence as shown by the table in Fig.~\ref{fig3a}(b) (for symmetric TSP) and Fig.~\ref{fig3b}(b) (for asymmetric TSP). The optimal Hamiltonian cycle $\mathcal{H}_b$ with the minimum $T$ is depicted on the Bloch sphere in Figs.~\ref{fig3a}(a)~and~\ref{fig3b}(a), which is found using a brute-force approach.
This result is proof of principle for our encoding scheme on the Bloch sphere. 
To solve these 5-city TSPs using a quantum approach, quantum parallelism is used on the Bloch sphere, shown in Figs.~\ref{fig3a}(c) and \ref{fig3b}(c). For optimizing the quantum algorithm, there are a few control knobs such as the initial state, strength of the rotation operators, and number of iterations in SPSA. The choice of the initial city $\ket{g_0}$ affects the population of the quantum states throughout the protocol. Different initial cities favor different sets of TSP paths, thus changing them increases the probability of finding the optimal one. The colored paths in Figs.~\ref{fig3a}(c) and \ref{fig3b}(c) with varying opacity show the population transfer between the quantum states. The random seed of the optimizer provides different initializations for the rotation operators, resulting in exploring more paths, one of which may be the optimal one. In this case, $5$ (out of $44$ $U^u$) intermediate rotation operators are chosen such that there is at least one \textit{up} operator $U^u$ between layers guiding the population transfer for an optimal outcome of the algorithm. All other operators ($39$ and $44$ different $U^u$ and $U^d$ respectively) are kept constant during the optimization process. Adding more numbers of $U^u$ along with \textit{down} operators $U^d$ can help while solving a more complex problem. Increasing the hyper-parameters, such as the hopping step and the total number of iterations in SPSA, also helps to converge to the optimal solution as more optimization landscape is covered.

\begin{figure}[t]
\centering
\includegraphics[scale = 0.50,trim={6.45cm 8cm 8cm 0cm},clip]{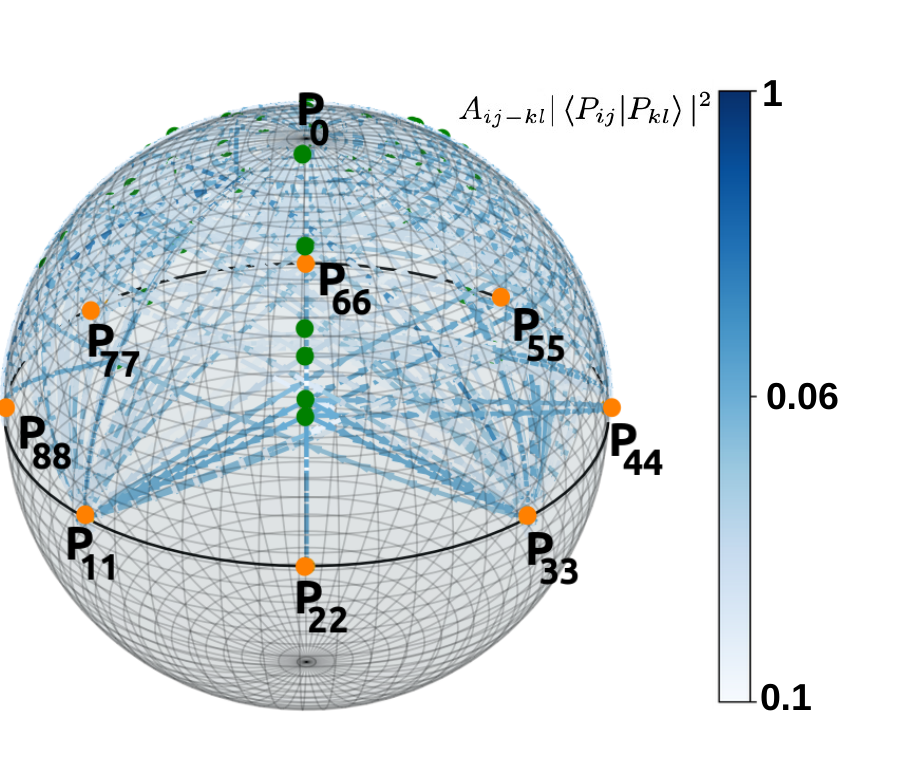}
\caption{Bloch sphere representing multiple paths for a symmetric $8$-city TSP (given by Eq.~(\ref{cm8}) in appendix \ref{costm}) for the optimal parameters which are found using the quantum approach discussed in Section \textbf{\RNum{3}}(b). The dominant route is found to be $P_{22}\rightarrow$ $P_{28} \rightarrow$ $ P_{88} \rightarrow$ $ P_{86} \rightarrow$ $ P_{66} \rightarrow$ $ P_{64} \rightarrow$ $ P_{44} \rightarrow$ $ P_{47} \rightarrow$ $ P_{77} \rightarrow$ $ P_{75} \rightarrow$ $ P_{55} \rightarrow$ $ P_{51} \rightarrow$ $ P_{11} \rightarrow$ $ P_{12} \rightarrow P_{22}$ which correspond to the classical optimal TSP path $c_{2}\rightarrow$ $c_{8}\rightarrow$ $ c_{6}\rightarrow$ $ c_{4}\rightarrow$ $ c_{3} \rightarrow$ $ c_{7}\rightarrow$ $ c_{5}\rightarrow$ $ c_{1}\rightarrow c_{2}$.
}
\label{fig5a}
\end{figure}

Our algorithm can efficiently and accurately solve TSP with a higher number of cities as shown in Fig.~\ref{fig5a}. An $8$-city problem (cost matrix given by Eq.~(\ref{cm8}) in appendix \ref{costm}) is solved on a single qubit, where we see that the superposition of paths (blue paths) gets fairly complex, requiring $12$ operators $U^u$ to be varied. The complexity is reflected by numerically observing many local minima in the optimization landscape for a larger number of cities. In such a case, the SPSA algorithm is best suited to find the optimal parameters compared to gradient-and simplex-based algorithms. The stochasticity of SPSA assists in navigating the optimization landscape more efficiently by overcoming the issue of getting stuck in local minima. The global convergence of the SPSA algorithm can be studied in detail \cite{he2003convergence}, which is currently beyond the scope of this work. 

\begin{figure}[t]
\includegraphics[scale = 0.40,trim={4cm 6cm 5cm 0cm},clip]{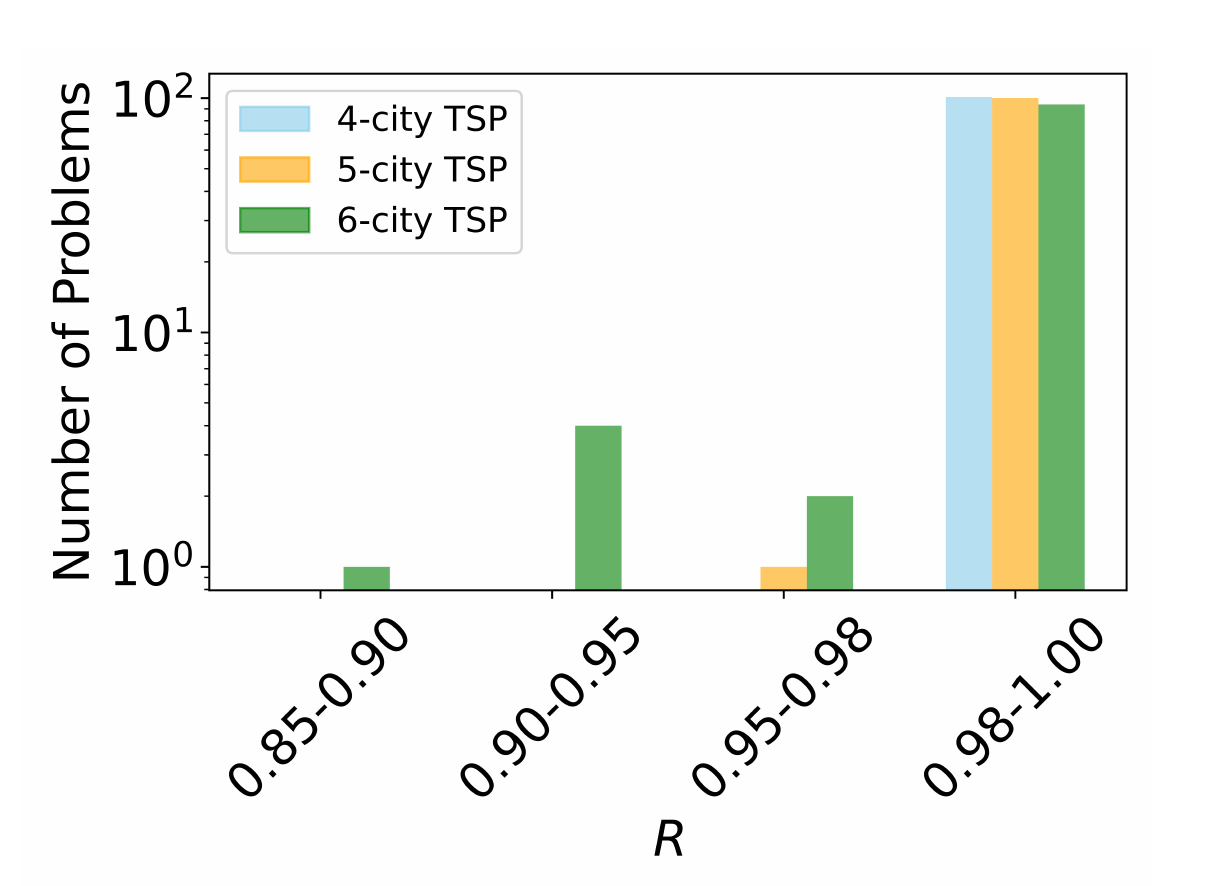}
\caption{Histogram showing the approximation ratio $R$ and the number of problems solved out of a hundred instances of 4-city, 5-city, and 6-city TSP using the SPSA optimizer.}
\label{fig5}
\end{figure}

Fig.~\ref{fig5} shows the robustness of our encoding and the protocol when solving large instances of different TSP problems. The results for $100$ randomly generated cost matrices each for $4$, $5$, and $6$ city TSPs are summarized. 
The histogram shows that all instances of $4$-city TSP are successfully solved while a majority i.e. more than $98\%$ of $5$-city and more than $90\%$ of $6$-city problems, are successfully tackled. 
For the instances that are not solved exactly by our algorithm, we still get solutions with a high ($R\geq 0.90$) approximation ratio, where $R = D_{min}/D_{ob}$ measures the optimality of the obtained solution $D_{ob}$ as compared to the global solution $D_{min}$. 
One of the ways to increase the number of solved problems is by changing the choice of the initial city as well as the initial superposition state. Furthermore, adding a larger number of varying rotation operators responsible for the population transfer of states to the optimizing parameters also helps. Finally, when the number of iterations of SPSA increases from $1000$ to $5000$ along with $6$ sets of different $U^u$, the algorithm consistently solves more than $95\%$ of the problems exactly.

\begin{figure}[t]
\centering
\includegraphics[width = 0.45\textwidth,trim={17.5cm 0cm 10cm 0cm},clip]{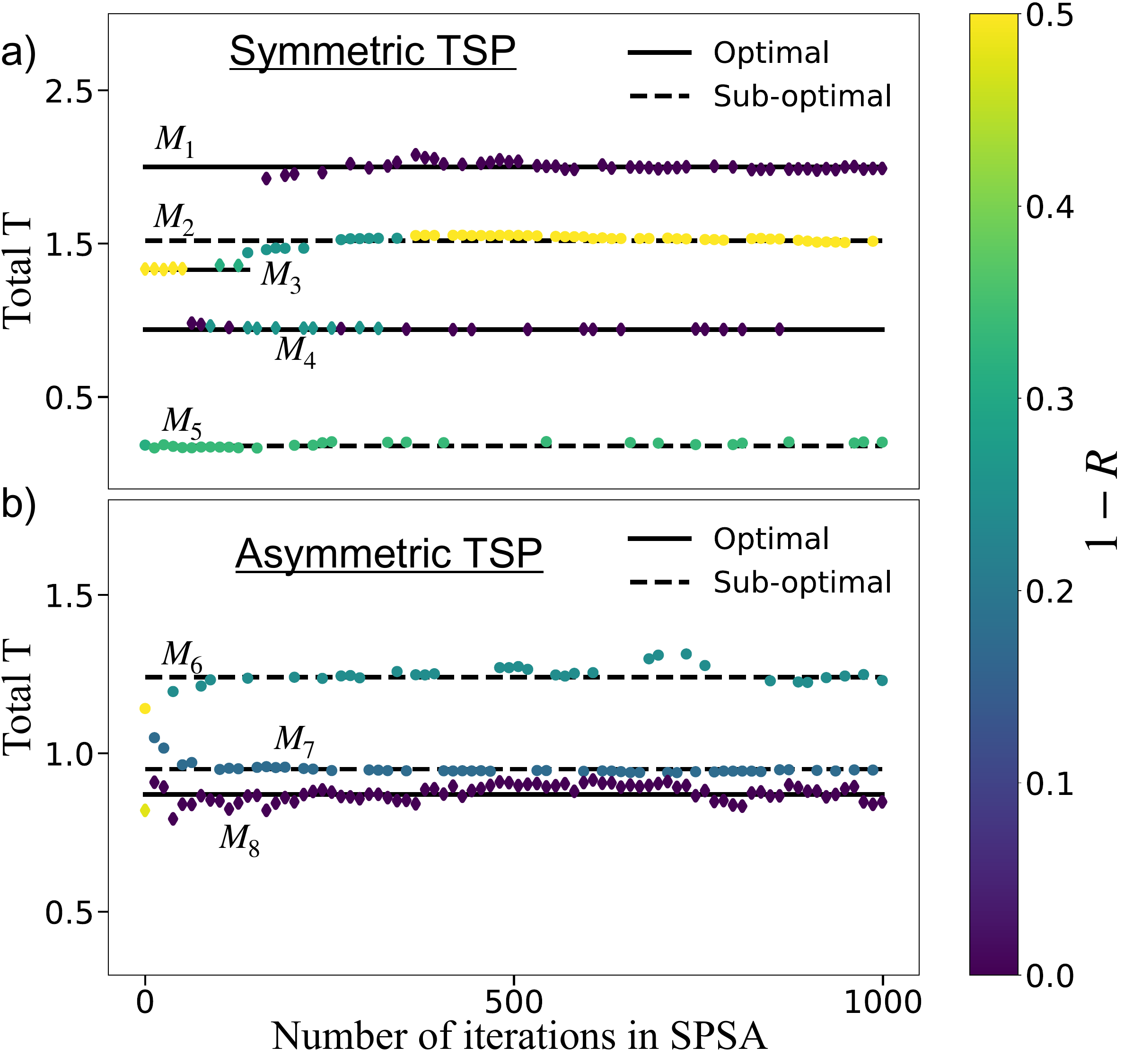}
\caption{Optimization landscape of a $5$-city TSP characterized by the parameter $T$ which is plotted against the number of iterations of SPSA, the color of the markers indicate the approximation ratio error ($1-R$). The plateaus (indicated by circles with a schematic background of dashed lines) in (a) $M_2,M_5$ for a symmetric and (b) $M_6,M_7$ for an asymmetric problem, are obtained using $1000$ iterations of SPSA that converges to sub-optimal solutions. The optimal solutions for both problems reside in the sub-spaces formed by the diamonds where the plateaus are shown with solid lines ($M_1,M_3,M_4,M_8$), which are found by optimizing the hyper-parameters and, thereby exploring other plateaus.}
\label{fig6}
\end{figure}

\begin{figure}[t]
\includegraphics[width = 0.45\textwidth,trim={1cm 0cm 0cm 0cm},clip]{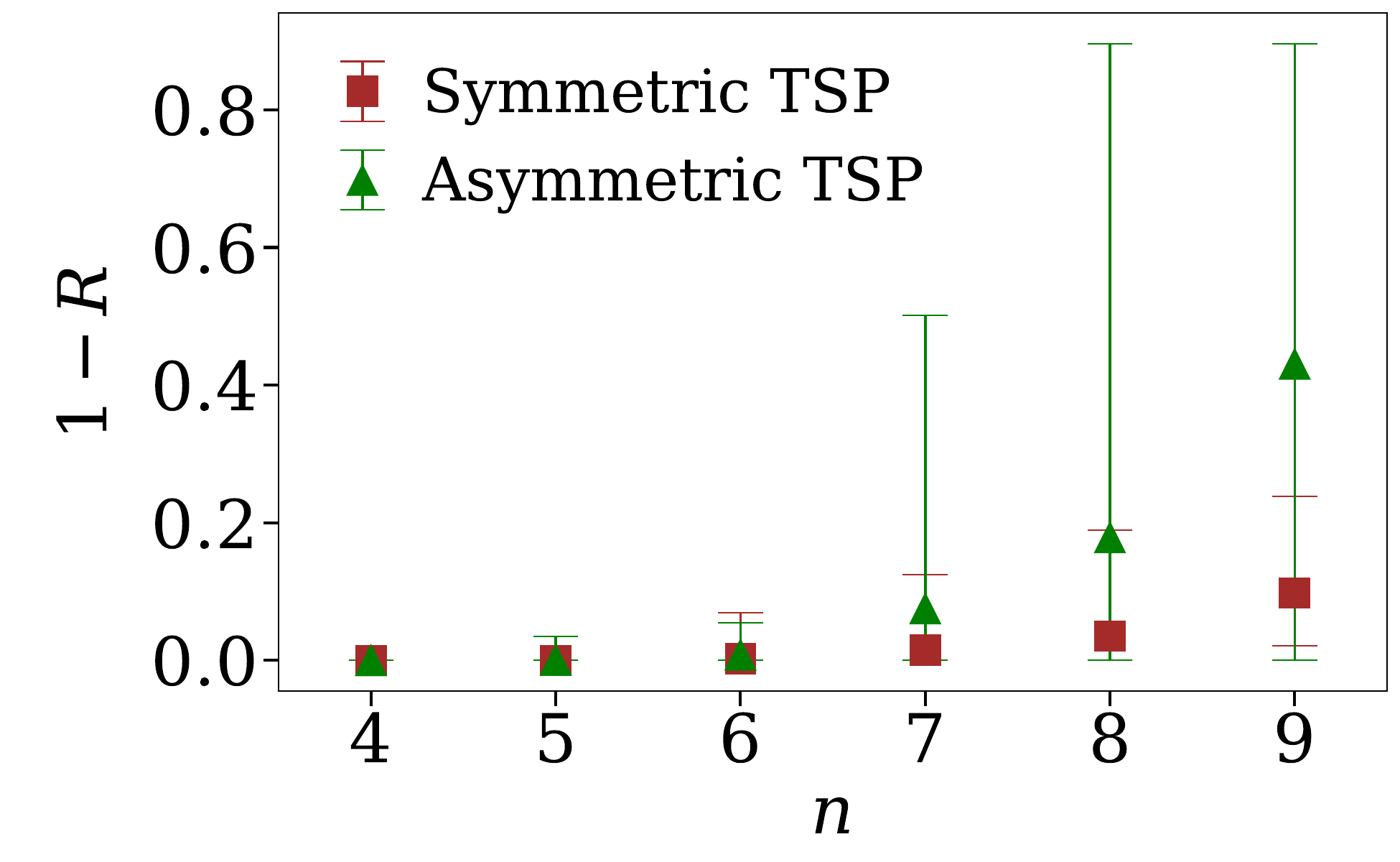}
\caption{Approximation ratio error ($1-R$) in finding the TSP solution for $100$ different instances of $n= 4-9$ city symmetric/asymmetric problems, where green triangles and brown squares correspond to the mean value of $1-R$ that increases with problem size. During optimization, a maximum $0.1\%$ random noise is added to the rotation operators accounting for the experimental imperfection, leading to an increase in the errors (indicated by the error bars).}
\label{fig7}
\end{figure}

Some insight is now provided into the complexity of the TSP control landscape and why, in certain rare cases, the exact solution is not obtained.
As explained earlier in Section \textbf{\RNum{3}(b)}, $T$ (Eq.~(\ref{T})) cannot be used as the objective function for the quantum version of the algorithm. Nevertheless, $T$ is useful to characterize the high-dimensional optimization landscape as it hints towards potential barren plateaus.
Fig.~\ref{fig6} shows the qualitative behavior of $T$ plotted against the number of iterations for SPSA, with the colored points indicating the error in the approximation ratio ($1-R$).
The parameters for $5$-city symmetric (panel (a)) and asymmetric (panel (b)) problems are found by the SPSA optimizer with $1000$ iterations. The obtained results lie either on the schematic dashed lines $M_2,M_5,M_6,M_7$ or solid lines $M_1,M_3,M_4,M_8$ corresponding to the manifolds containing sub-optimal or optimal solutions respectively. 
The sub-optimality is characterized by a higher value of $1-R$ (shown by the color bar) that occurs for a particular set of initialized hyper-parameters which also corresponds to a two-fold degeneracy or near-degeneracy ($M_2,M_5$ in (a) and $M_6,M_7$ in (b)) in the values of $T$. 
In the case of sub-optimal dynamics, the optimizer seems to be stuck between two $T$ regimes where the obtained results fail to converge to the global minima even if the number of iterations is increased.
However, the sub-optimal solutions (indicated as circles) are improved to optimal (indicated as diamonds) by training the hyper-parameters of the optimizer, as shown in Figs.~\ref{fig6}(a,b).
Specifically, tuning hyper-parameters such as the initial city, the choice of $U^u$, and the hopping step in SPSA allows the optimizer to jump to other $T$ manifolds where the optimal solution resides ($M_1,M_3,M_4$ in (a) and $M_8$ in (b)). Additionally, the observed flatness in $T$ can assist in reducing the number of iterations required to solve the problem. By changing the random seed early in the process (during the initial few iterations), the optimizer can explore other sub-spaces of the parameters to find the solution.

Using example TSPs, we now apply the quantum optimal control to a noisy system, simulating conditions closer to a real experiment. 
Fig.~\ref{fig7} displays the approximation ratio error $1-R$ for four- to nine-city symmetric and asymmetric TSPs with a maximum of $0.1\%$ random error (typical in experiments) to the rotation operators \cite{PhysRevLett.117.060504}.
For each $n$-city problem, $10$ different cost matrices are considered each of which is solved for $10$ instances of random errors, making up $100$ samples. The average $1-R$ (indicated by the green triangles and brown squares) for a particular $n$-city TSP scales with $n$ for both the cases, due to an expected increase in complexity. This overall trend of $1-R$ is attributed to the complexity of optimizing a larger number of $U^u$ ($44$ for $5$-city, and $408$ for $9$-city problems) for increasing $n$, akin to increasing depth in variational algorithms \cite{farhi2014quantum,peruzzo_variational_2014}.
The error bars in the $1-R$ also widen with $n$, indicating a higher sensitivity to the optimal parameters.
With increasing problem size $n$, more manifolds containing sub-optimal solutions emerge. This can facilitate easier switching between the manifolds containing optimal and sub-optimal solutions (as shown by Fig.~\ref{fig6}).
The larger the sample size, the probability of obtaining a pathological problem that gets stuck at sub-optimal manifolds increases, contributing to larger error bars. The asymmetric TSPs are harder to solve as compared to symmetric TSPs, exhibiting higher values of $1-R$ and larger error bars, which is suggestive of more multiple local minima for this case. 
Another indicator that can help us to understand the trend in Fig.~\ref{fig7} is the percentage of problems solved with $R \geq 0.99$ (as shown in Fig.~\ref{fig5}). Numerically we observe that this percentage decreases with problem size, specifically $\sim 97\%$ ($ 81\%$) for $6$-city symmetric (asymmetric) and $\sim 62\%$ ($38\%$) for $7$-city symmetric (asymmetric) problems, which is consistent with a general increase in complexity from $6$-city to $7$-city problems.

\section{\label{conc}Discussion and Outlook}
Combinatorial problems such as the TSP have been extensively mapped to many real-world optimization problems. However, solving them on a quantum computer with minimum resources and high accuracy is a big challenge. 
Current quantum algorithms rely upon mapping the problem either to a QUBO form or using a gate-based approach, both of which suffer from the requirement of a large number of noiseless qubits. 
Keeping the ambitious goal of efficiently tackling \textit{real-world} TSPs using quantum systems, we introduce a scheme that maps the problem to a single-qubit Bloch sphere and uses quantum parallelism to solve it.
Algorithms implemented on one qubit exploit the controllability in creating a selective superposition of states and offer a polynomial speed-up over the classical counterpart. 

Our scheme will act as a template to further develop algorithms that utilize the superposition principle for resource efficiency and extend this to entangled qubits for gaining a quantum advantage in the NISQ era. It can be implemented on any quantum platform that allows control of rotating a qubit along any direction by arbitrary values with high fidelity.
We map the cities of a TSP to the quantum states on the Bloch sphere which are traversed by optimally selected rotation operators.
The embedding of a $n$-city problem is such that there are $n$ quantum states on the equator of the Bloch sphere each connected to one of the poles through a geodesic path each containing $n-1$ quantum states to encode the relative distances. 
All the possible routes for a given problem are accessed by creating a superposition of the quantum states at each geodesic and the optimal route is found by calibrating the rotation operators using the SPSA optimizer. We show that for four- to six-city TSP, the algorithm finds the exact solution for most of the problem instances, which is much better than the current quantum schemes in terms of accuracy and resources. The protocol becomes more sensitive to noise/errors when more quantum states are placed closely on a geodesic. 

There is a notable similarity between the quantum superposition construction used for TSP and Grover's algorithm for unstructured search \cite{10.1145/237814.237866}. Similar to Grover’s algorithm, it was demonstrated in \cite{lloyd1999quantum}, that quantum search algorithms based solely on the superposition principle can achieve the polynomial speedup ($O(\sqrt{N})$) over classical algorithms. Finding the optimal TSP path among the various allowed paths on a Bloch sphere can be treated as a quantum search problem, and the same polynomial speedup over classical algorithms seen in \cite{lloyd1999quantum} can be expected from our algorithm. However, in order to rigorously establish a polynomial quantum advantage for our algorithm, we need to define concepts such as ``quantum phase oracle'' and ``inversion about the average'' which will be the focus of future work. 

Furthermore, geometric visualization of quantum problems can be a powerful tool in itself \cite{PhysRevA.49.4101,PhysRevA.91.042122,PhysRevA.97.043606}. Our work also provides a unique geometrical visualization of how a quantum algorithm tackles the TSP problem on a Bloch sphere. Such visualization can prove to be useful in gaining insights into the future development of algorithms and finding patterns for the optimal set of rotation operators.

\begin{acknowledgments}
This work is funded by the German Federal Ministry of Education and Research within the funding program “Quantum Technologies - from basic research to market” under Contract No. 13N16138.
\end{acknowledgments}

\appendix

\setcounter{equation}{0}
\setcounter{figure}{0}
\setcounter{table}{0}
\setcounter{section}{0}
\renewcommand{\thesection}{A}
\makeatletter
\renewcommand{\theequation}{A\arabic{equation}}
\renewcommand{\thefigure}{A\arabic{figure}}
%\renewcommand{\bibnumfmt}[1]{[S#1]}
%\renewcommand{\citenumfont}[1]{S#1}
%%%%%%%%%% Prefix a "S" to all equations, figures, tables and reset the counter %%%%%%%%%%

\section{Cost matrices for the example $n$-city TSPs}
\label{costm}
The cost matrices corresponding to the prototypical TSPs solved in the main text are provided in this section.
For the example symmetric $4$-city TSP used to generate Figs.~\ref{fig2}(b) and \ref{fig4}(b), the cost matrix is given as
\begin{align}
B^{4S}_{ij}=\begin{bmatrix} \label{cm4}
0 & 0.57 & 0.78 & 0.55 \\
0.57 & 0 & 0.41 & 0.90 \\
0.78 & 0.41 & 0 & 0.30  \\
0.55 & 0.90 & 0.30 & 0
\end{bmatrix}.
\end{align}

\noindent The results section show solutions to the following symmetric $B^{5S}_{ij}$ and asymmetric $B^{5A}_{ij}$ $5$-city TSP in Figs.~\ref{fig3a}~and~\ref{fig3b}, respectively. 
\begin{align}
B^{5S}_{ij}=\begin{bmatrix} \label{cm1}
0 & 0.35 & 0.48 & 0.28 & 0.63 \\
0.35 & 0 & 0.28 & 0.13 & 0.31\\
0.48 & 0.28 & 0 & 0.45 & 0.78 \\
0.28 & 0.13 & 0.45 & 0& 0.33 \\
0.63 & 0.31 & 0.78 & 0.33 & 0\\
\end{bmatrix}, 
\end{align}
\begin{align}
B^{5A}_{ij} = \begin{bmatrix} \label{cm2}
0 & 0.58 & 0.92 & 0.20 & 0.97 \\
0.44 & 0 & 0.98 & 0.83 & 0.06\\
0.92 & 0.19 & 0 & 0.54 & 0.09 \\
0.02 & 0.14 & 0.15 & 0& 0.98 \\
0.94 & 0.34 & 0.05 & 0.80 & 0\\
\end{bmatrix}.
\end{align}

\noindent Fig.~\ref{fig5a} provides the optimal route of a symmetric $8$-city TSP whose cost function is given as,
\begin{align}
B^{8S}_{ij}  = \begin{bmatrix} \label{cm8}
0 & 0.84 & 0.46 & 0.60 & 0.36 & 0.67 & 0.40 & 0.40 \\
0.84 & 0 & 0.59 & 0.36 & 0.49 & 0.28 & 0.58 & 0.20 \\
0.46 & 0.59 & 0 & 0.41 & 0.81 & 0.15 & 0.64 & 0.28 \\
0.60 & 0.36 & 0.41 & 0 & 0.73 & 0.27 & 0.42 & 0.47 \\
0.36 & 0.49 & 0.81 & 0.73 & 0 & 0.20 & 0.26 & 0.59 \\
0.67 & 0.28 & 0.15 & 0.27 & 0.20 & 0 & 0.53 & 0.58 \\
0.40 & 0.58 & 0.64 & 0.42 & 0.26 & 0.53 & 0 & 0.76 \\
0.40 & 0.20 & 0.28 & 0.47 & 0.59 & 0.58 & 0.76 & 0
\end{bmatrix}
\end{align}

\renewcommand{\thesection}{B}
\renewcommand{\theequation}{B\arabic{equation}}
\renewcommand{\thefigure}{B\arabic{figure}}

\section{Calculation of the number of paths and operators}
\label{calc}

The calculation for finding the total number of allowed paths, the total number of paths in the quantum algorithms, the number of forbidden paths, and the number of independent rotational operators are discussed here. 

\subsection{Allowed paths in classical TSP}
The number of allowed paths is the number of classical Hamiltonian cycles for a given TSP. 
For a $n$-city TSP, the number of Hamiltonian cycles $N_H$ is given by
\begin{equation}
    N_{H} = (n-1)!
\end{equation}
which is the total number of cyclic permutations, also shown in Fig.~\ref{fig2}.

\subsection{Total number of paths in the quantum algorithm}
The total number of paths in the quantum algorithm is different from the allowed path calculated above. In the quantum algorithm, there are many more possible paths due to simultaneously traversing on the Bloch sphere. This is also depicted in Fig.~\ref{fig4} which has more paths as compared to the routing chart given in Fig.~\ref{fig2}. By construction, there are multiple layers $L_i$ in the routing chart, each containing a vertical array of rectangles (or circles) $P_{ij}$ representing the cities placed on the Bloch sphere. To construct a path in the routing chart, one rectangle (or circle) from each layer has to be selected and connected through the arrows.

\begin{itemize}  
    \item The paths going from layer $L_i$ to $L_{i+1}$ are found by connecting all of the rectangles (or circles) of $L_i$ to all the rectangles (or circles) of the subsequent layer $L_{i+1}$. Multiplying the number of rectangles (or circles) of each layer gives the total number of paths possible between $L_i$ and $L_{i+1}$. 
    
    \item Similarly, the next part of the route is constructed by connecting the rectangles (or circles) of $L_{i+1}$ to $L_{i+2}$. Hence, the total number of possible routes is given by multiplying the number of rectangles (or circles) of each subsequent layer for the whole routing chart. Now for an $n$-city problem, there are a total of $2n+1$ layers. We divide the layers into two categories, the edge layers and the middle layers.
    
    \item Edge layers $L_0,L_1,L_{2n-2},L_{2n-1},L_{2n}$: The first $L_0$ and the last $L_{2n}$ layers have only one rectangle (or circle), while the $L_1$, $L_{2n-1}$ and $L_{2n-2}$ have $n-1$ rectangles (or circles) each by construction. So, from these layers, the contribution to the total number of paths is,
    \begin{equation}
        1(n-1)(n-1)(n-1)1 = (n-1)^3
    \end{equation}
    \item Middle layers: The rest of the $2n-4$ layers are left in the middle of the routing chart, where the construction is such that a block of two consecutive layers repeats itself $n-2$ times. This block has one layer with $n-1$ rectangles (or circles) and the next one fanning out to $(n-1)(n-2)$ rectangles (or circles), which contributes $(n-1)(n-1)(n-2) = (n-1)^2(n-2)$ to the calculation of the total paths. As mentioned, there are $n-2$ of such blocks, which makes a total contribution,
    \begin{equation}
        [(n-1)^2(n-2)]^{n-2}
    \end{equation} 
    of the middle layer rectangles (or circles). 
\end{itemize}

Hence, the total number of possible paths in the quantum case is then given by multiplying the contribution of the middle blocks and the edge layers, which is given as,
    \begin{equation}
        N_T = (n-1)^3[(n-1)^2(n-2)]^{n-2}
    \end{equation}

\subsection{Number of forbidden paths}
\label{forbnums}
To find out the number of forbidden paths, first, the total number of paths in the quantum algorithm is calculated above from the routing chart. 
The total number of forbidden paths $N_F$ is the difference between the total number of possible paths $N_T$ and the total number of allowed paths $N_H$, 
\begin{equation}
\begin{split}
\begin{aligned}
  N_F &= N_T- N_H \\
  &= (n-1)^3[(n-1)^2(n-2)]^{n-2} - (n-1)!   
\end{aligned}
\end{split}
 \label{forbnum}
\end{equation}

\subsection{Number of rotational operators}
\label{rotnum}
The routing chart is used here again to calculate the total number of rotational operators. For this, we consider all the tunable operators, which are constructed by taking into account the classical routes, shown by the arrows in Fig.~\ref{fig2}. 
The number of arrows in the routing chart corresponding to all the allowed paths provides the number of rotational operators, which is counted below. The layers are divided again into edge layers ($L_0,L_1,L_{2n},L_{2n-1},L_{2n-2}$) and the middle layers with $n-2$ blocks each containing two layers.
\begin{itemize}
    \item Edge layers: The number of arrows that are originating from the edge layers is found to be $4(n-1)$, each corresponding to an independent rotational operator. 
    \item Middle layers: For each block in the middle layers, there are $2(n-1)(n-2)$ arrows originating within that block. So, for a total of $n-2$ middle blocks, we get $2(n-1)(n-2)(n-2) = 2(n-1)(n-2)^2$ arrows. 
\end{itemize}
The total number of arrows and hence the total number of independent rotational operators are given by adding the numbers of both cases, that is,
\begin{equation}
\begin{split}
\begin{aligned}
  N_O &= 4(n-1)+2(n-1)(n-2)^2 \\
  &= 2(n-1)(2+(n-2)^2) 
\end{aligned}
\end{split}
 \label{opernum}
\end{equation}

\bibliographystyle{unsrt}
\bibliography{main.bib}

\end{document}